\documentclass{aipproc}

\layoutstyle{6x9}

\usepackage{amsmath}
\usepackage{amssymb}
\usepackage{mathbbol}
\usepackage{mathptm}
\usepackage{times}

\begin{document}

%%%%%%%%%%%%%%%%%%%%%%%%%%%%%%%%%%%%%%%%%%%%%%%%%%%%%%%%%%%%%%%%%%%%%%
%                                                                    %
% Title                                                              %
%                                                                    %
%%%%%%%%%%%%%%%%%%%%%%%%%%%%%%%%%%%%%%%%%%%%%%%%%%%%%%%%%%%%%%%%%%%%%%

\title{Azimuthal single-spin asymmetries in semi-inclusive
deep-inelastic scattering on a transversely polarised hydrogen target}

\author{Markus Diefenthaler (on behalf of the \textsc{HERMES}
collaboration)} 
{address={Physikalisches Institut II,
Friedrich-Alexander-Universit\"at Erlangen-N\"urnberg,
Erwin-Rommel-Stra{\ss}e 1, 91058 Erlangen, Germany},
email={markus.diefenthaler@desy.de}}

\begin{abstract}
Azimuthal single-spin asymmetries (\textsc{SSA}) in semi-inclusive
electroproduction of charged pions and kaons in deep-inelastic
scattering of positrons on a transversely polarised hydrogen target
were observed. \textsc{SSA} amplitudes for both the Collins and the
Sivers mechanism are presented.
\end{abstract}


%%%%%%%%%%%%%%%%%%%%%%%%%%%%%%%%%%%%%%%%%%%%%%%%%%%%%%%%%%%%%%%%%%%%%%
%                                                                    %
% Physics and Astronomy Classification Scheme (PACS 2003):           %
%                                                                    %
% 13.60.-r	Photon and charged-lepton interactions with hadrons  %
%               (for neutrino interactions, see 13.15.+g)            %
% 13.88.+e 	Polarisation in interactions and scattering          %
% 14.20.Dh 	Protons and neutrons                                 %
% 14.65.-q 	Quarks                                               %
%                                                                    %
%%%%%%%%%%%%%%%%%%%%%%%%%%%%%%%%%%%%%%%%%%%%%%%%%%%%%%%%%%%%%%%%%%%%%%

\maketitle

In 2005 the \textsc{HERMES} collaboration published first evidence for
azimuthal single-spin asymmetries (\textsc{SSA}) in the semi-inclusive
production of charged pions on a transversely polarised target
\cite{Airapetian:2004tw}. Significant signals for both the Collins and
Sivers mechanisms were observed in data recorded during the 2002--2003
running period of the \textsc{HERMES} experiment.  Below we present a
preliminary analysis of these data combined with additional data taken
in the years 2003 and 2004. All data were recorded at a beam energy of
\(27.6\,\text{\texttt{GeV}}\) using a transversely nuclear-polarised
hydrogen-target internal to the \textsc{HERA} positron storage ring at
\textsc{DESY}. The \textsc{HERMES} dual radiator ring-imaging
\v{C}erenkov counter allows full \(\pi^{\pm}\), \(K^{\pm}\), \(p\)
separation for all selected particle momenta. Therefore, a preliminary
analysis of \textsc{SSA} in the electroproduction of charged kaons
on a transversely polarised target is also presented.

At leading twist, the momentum and spin of the quarks inside the
nucleon are described by three parton distribution functions: the
well-known momentum distribution \(q\left(x,Q^{\,2}\right)\), the
known helicity distribution \(\Delta\,q\left(x,Q^{\,2}\right)\)
\cite{Airapetian:2004zf} and the unknown \emph{transversity
distribution} \(\delta\,q\left(x,Q^{\,2}\right)\)
\cite{Ralston:1979ys, Artru:1989zv, Jaffe:1991kp, Cortes:1991ja}. In
the helicity basis, transversity is related to a quark-nucleon forward
scattering amplitude involving helicity flip of both nucleon and quark
(\(N^{\Rightarrow}q^{\leftarrow} \boldsymbol{\rightarrow}
N^{\Leftarrow}q^{\rightarrow}\)). As it is chiral-odd, transversity
cannot be probed in inclusive measurements. At \textsc{HERMES}
transversity in conjunction with the chiral-odd Collins fragmentation
function \cite{Collins:1992kk} is accessible in \textsc{SSA} in
semi-inclusive DIS on a transversely polarised target (\emph{Collins
mechanism}). The Collins fragmentation function describes the
correlation between the transverse polarisation of the struck quark
and the transverse momentum \(\boldsymbol{P}_{\,\text{h}\perp}\) of
the hadron produced. As it is also odd under naive time reversal
(T-odd) it can produce a \emph{\textsc{SSA}}, i.e.~a left-right
asymmetry in the momentum distribution of the produced hadrons in the
directions transverse to the nucleon spin \cite{Burkardt:2003yg}.

The \emph{Sivers mechanism} can also cause a \textsc{SSA}: The T-odd
Sivers distribution function \cite{Sivers:1989cc} describes the
correlation between the transverse polarisation of the nucleon and the
transverse momentum \(\boldsymbol{p}_{\perp}\) of the quarks within. A
non-zero Sivers mechanism provides a non-zero Compton amplitude
involving nucleon helicity flip without quark helicity flip
(\(N^{\Rightarrow}q^{\leftarrow} \boldsymbol{\rightarrow}
N^{\Leftarrow}q^{\leftarrow}\)), which must therefore involve orbital
angular momentum of the quark inside the nucleon
\cite{Burkardt:2003yg, Brodsky:2002cx}.

With a transversely polarised target, the azimuthal angle
\(\phi_{\,\text{S}}\) of the target spin direction in the
``\(\Uparrow\)'' state is observable in addition to the azimuthal
angle \(\phi\) of the detected hadron.  Both azimuthal angles are
defined about the virtual-photon direction with respect to the lepton
scattering plane.  The additional degree of freedom
\(\phi_{\,\text{S}}\), not available with a longitudinally polarised
target, results in distinctive signatures: \(\sin{\left(\phi +
\phi_{\,\text{S}}\right)}\) for the Collins mechanisms and
\(\sin{\left(\phi - \phi_{\,\text{S}}\right)}\) for the Sivers
mechanism \cite{Boer:1997nt}. Therefore, for all detected charged
pions and for each bin in \(x\), \(z\) or
\(\boldsymbol{P}_{\,\text{h}\perp}\) the cross section asymmetry for
unpolarised beam (U) and transversely polarised target (T) was
determined in the two dimensions \(\phi\) and \(\phi_{\,\text{S}}\):
\begin{equation*}
A_\text{UT}^{\pi^\pm}\left(\phi,\phi_{\,\text{S}}\right) = 
\frac{1}{\left|\,P_z\,\right|}\frac{
N_{\pi^\pm}^{\Uparrow}\left(\phi,\phi_{\,\text{S}}\right) -
N_{\pi^\pm}^{\Downarrow}\left(\phi,\phi_{\,\text{S}}\right)}
{N_{\pi^\pm}^{\Uparrow}\left(\phi,\phi_{\,\text{S}}\right) -
N_{\pi^\pm}^{\Downarrow}\left(\phi,\phi_{\,\text{S}}\right)}.
\end{equation*}
Here \(N_{\pi^\pm}^{\Uparrow\left(\Downarrow\right)}
\left(\phi,\phi_{\,\text{S}}\right)\) represents the semi-inclusive
normalised yield in the target spin state ``\(\Uparrow
\left(\Downarrow\right)\)'' and \(\left|\,P_z\,\right|=0.754\pm0.050\)
denotes the average degree of the target polarisation.

To avoid cross-contamination, the azimuthal amplitudes for the Collins
mechanism \(\left<\sin{\left(\phi +
\phi_{\,\text{S}}\right)}\right>_{\text{UT}}^{\pi^\pm}\) and the
Sivers mechanism \(\left<\sin{\left(\phi -
\phi_{\,\text{S}}\right)}\right>_{\text{UT}}^{\pi^\pm}\) were
extracted simultaneously. Recent studies showed that the terms for
\(\sin{\phi_{\,\text{S}}}\) and \(\sin{\left(2\phi-
\phi_{\,\text{S}}\right)}\) have to be added in the two-dimensional
least-squares fit for the asymmetry (the kinematic factors
\(A\left(\left<\,y\,\right>,\left<\,R\,\right>\right)\) and
\(B\left(\left<\,y\,\right>\right)\) are defined in
\cite{Airapetian:2004tw}):
\begin{eqnarray*}
A_\text{UT}^{\pi^\pm}\left(\phi,\phi_{\,\text{S}}\right) & = &
2\left<\sin{\left(\phi +
\phi_{\,\text{S}}\right)}\right>_{\text{UT}}^{\pi^\pm}
\frac{B\left(\left<\,y\,\right>\right)}{A\left(\left<\,y\,\right>,
\left<\,R\,\right>\right)}
\,\sin{\left(\phi+\phi_{\,\text{S}}\right)}+\\ &&2\left<\sin{\left(\phi
- \phi_{\,\text{S}}\right)}\right>_{\text{UT}}^{\pi^\pm}
\,\sin{\left(\phi-\phi_{\,\text{S}}\right)}+\\
&&2\left<\sin{\left(2\phi-\phi_{\,\text{S}}\right)}\right>_{\text{UT}}^{\pi^\pm}
\,\sin{\left(2\phi-\phi_{\,\text{S}}\right)} +
2\left<\sin{\phi_{\,\text{S}}}\right>_{\text{UT}}^{\pi^\pm}
\,\sin{\phi_{\,\text{S}}}.
\end{eqnarray*}
In  the  case  of charged  pions  least  squares is a  good  maximum
likelihood estimator.  However, in the  case of charged  kaons because
of the limited statistics maximum likelihood fits must be
used. Besides the  Collins and Sivers amplitude the likelihood
function is also maximised with respect to
\(\sin{\left(3\phi-\phi_{\,\text{S}}\right)}\),
\(\sin{\left(2\phi-\phi_{\,\text{S}}\right)}\) and
\(\sin{\phi_{\,\text{S}}}\) sine modulations.

In figure \ref{ssa-amplitudes} the virtual-photon Collins and Sivers
amplitudes as a function of \(x\), \(z\) and
\(\boldsymbol{P}_{\,\text{h}\perp}\) are plotted. The maximum
contribution of subleading longitudinal asymmetries to the
leading-twist Collins and Sivers amplitudes in figure
\ref{ssa-amplitudes} is \(0.004\), which is negligible compared to the
statistical uncertainties. In addition the simulated fraction of charged
pions originating from diffractive vector meson production and decay
is shown, to provide an estimate of the possible contribution from the
poorly known asymmetry of this process. The average values of the
kinematic variables in the experimental acceptance are
\(\left<\,x\,\right>=0.10\), \(\left<\,y\,\right>=0.53\),
\(\left<\,Q^{\,2}\,\right>=2.43\,\text{GeV}^{\,2}\),
\(\left<\,z\,\right>=0.36\),
\(\left<\,\boldsymbol{P}_{\,\text{h}\perp}\,\right>=0.40\,\text{GeV}\).
These preliminary results are based on nearly five times more
statistics than that in the publication \cite{Airapetian:2004tw} and
are consistent with the published result:

\begin{figure}
\begin{tabular}{cc}
\includegraphics[scale=0.38]{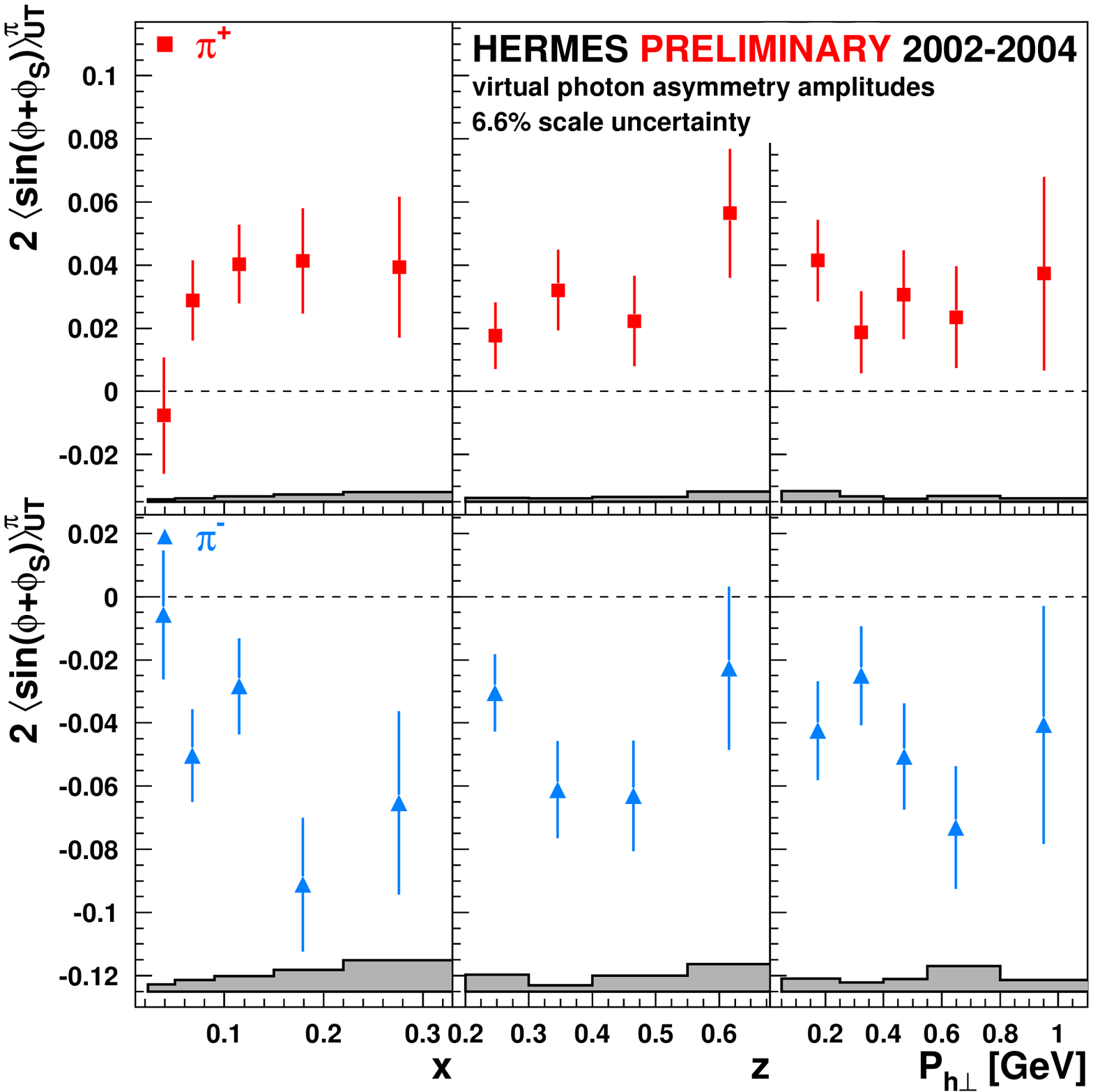} &
\includegraphics[scale=0.38]{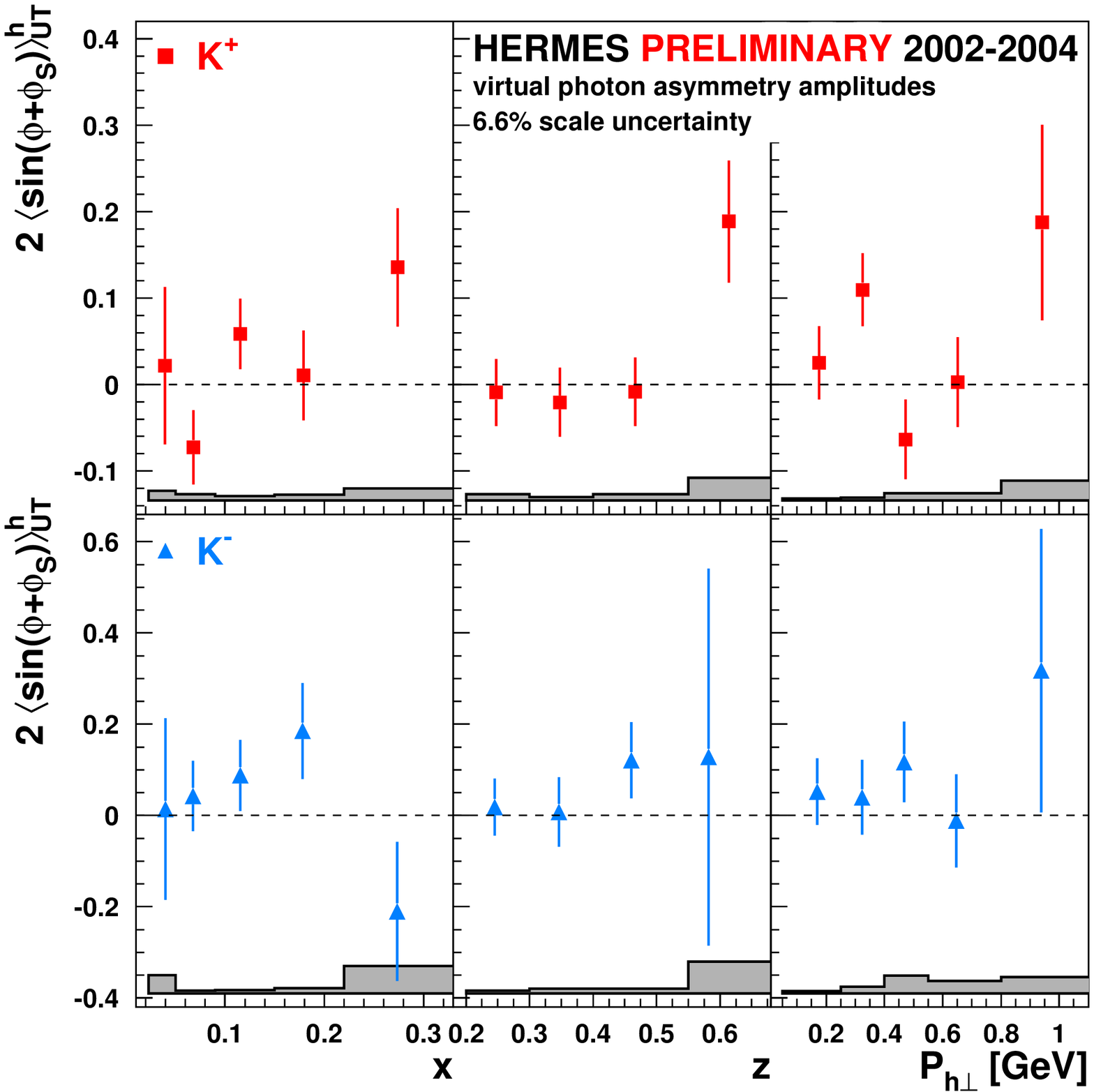} \\
\includegraphics[scale=0.38]{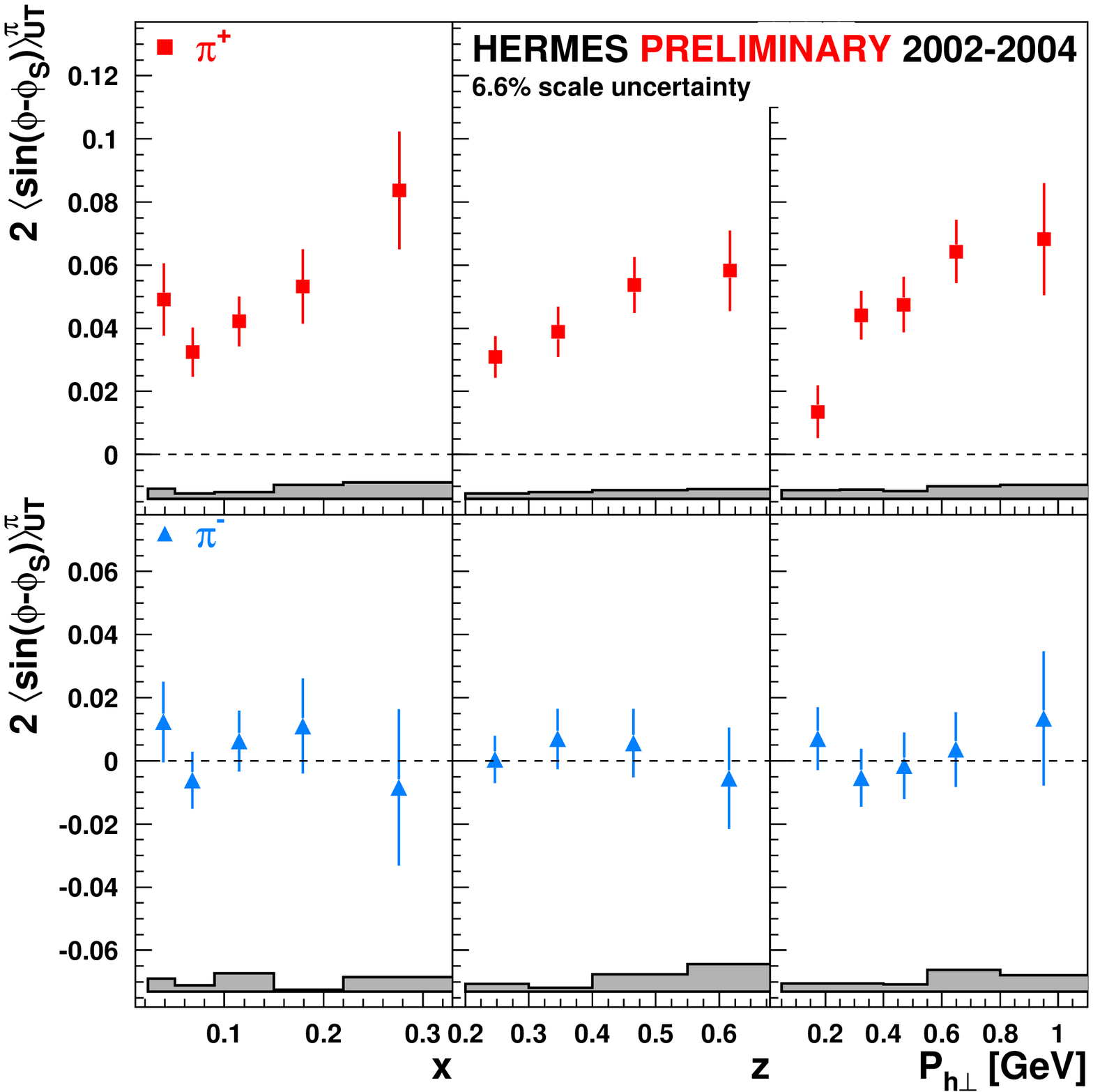} &
\includegraphics[scale=0.38]{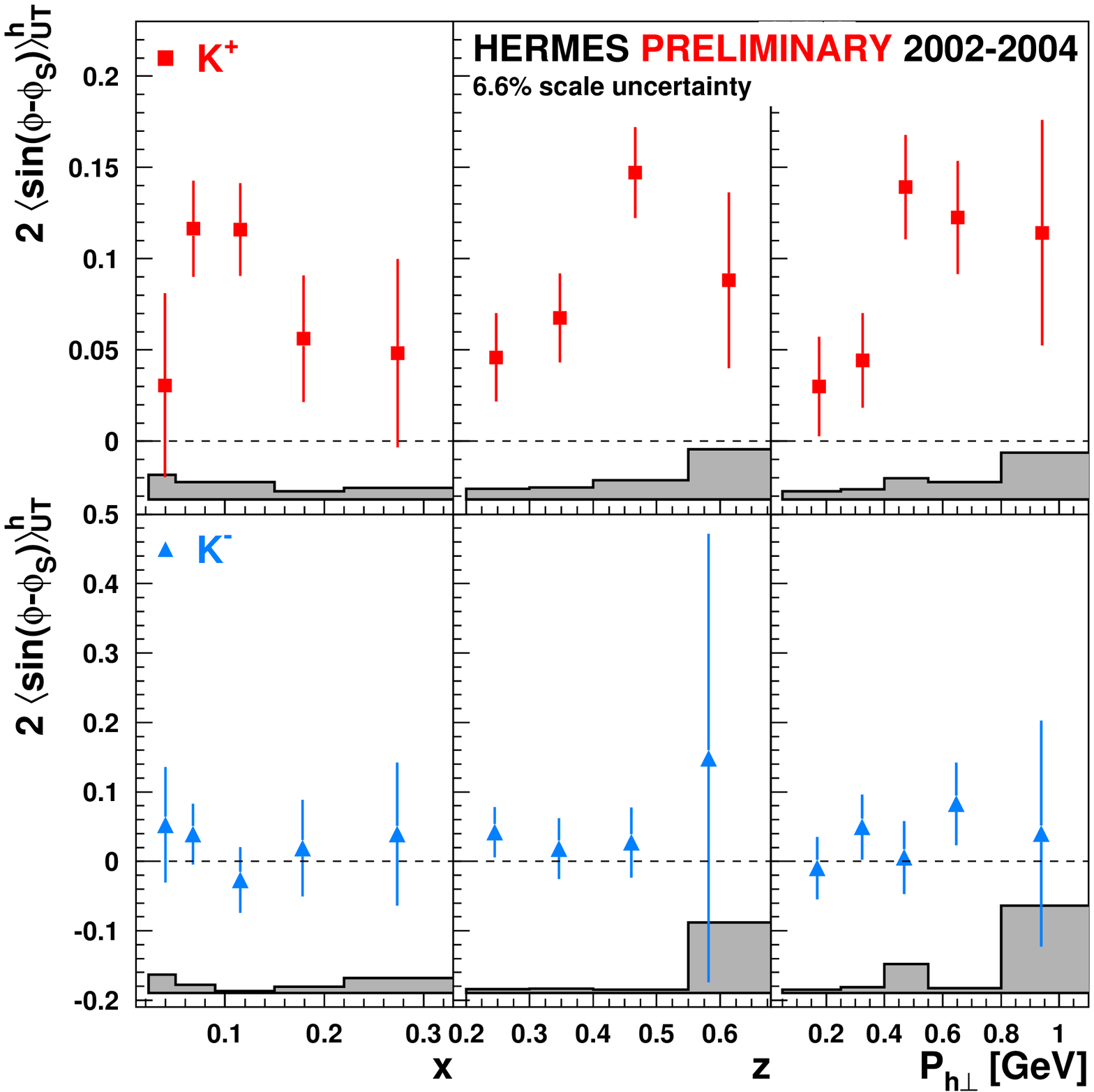}\\
\includegraphics[scale=0.38]{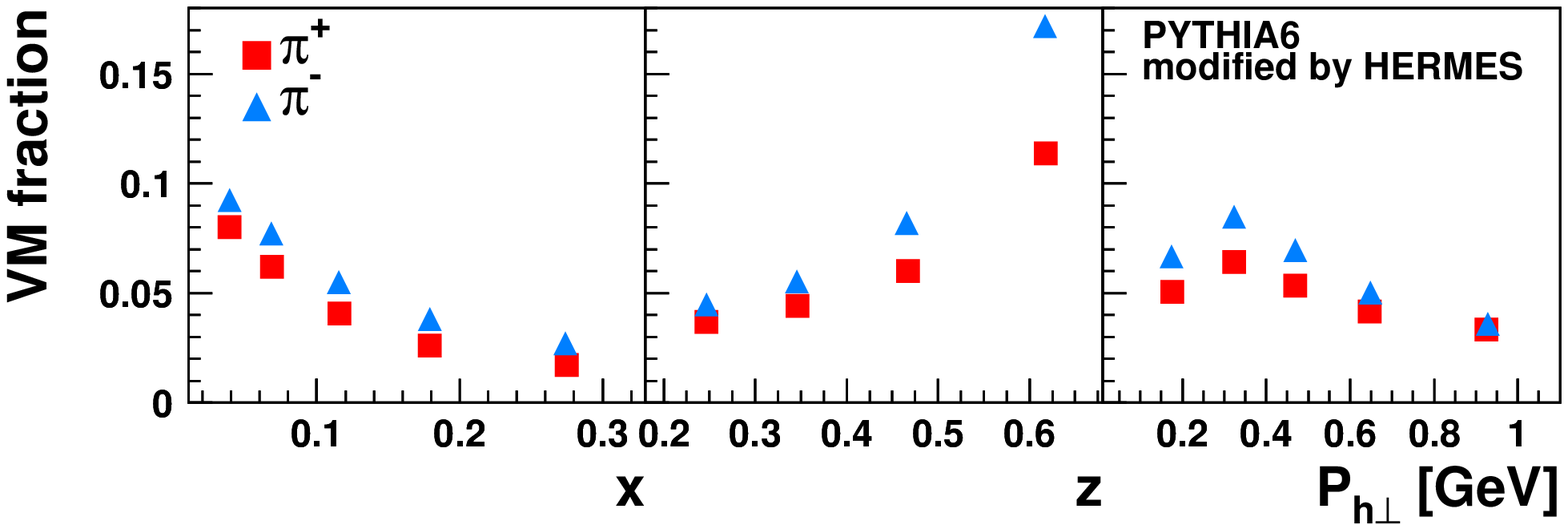} &
\includegraphics[scale=0.38]{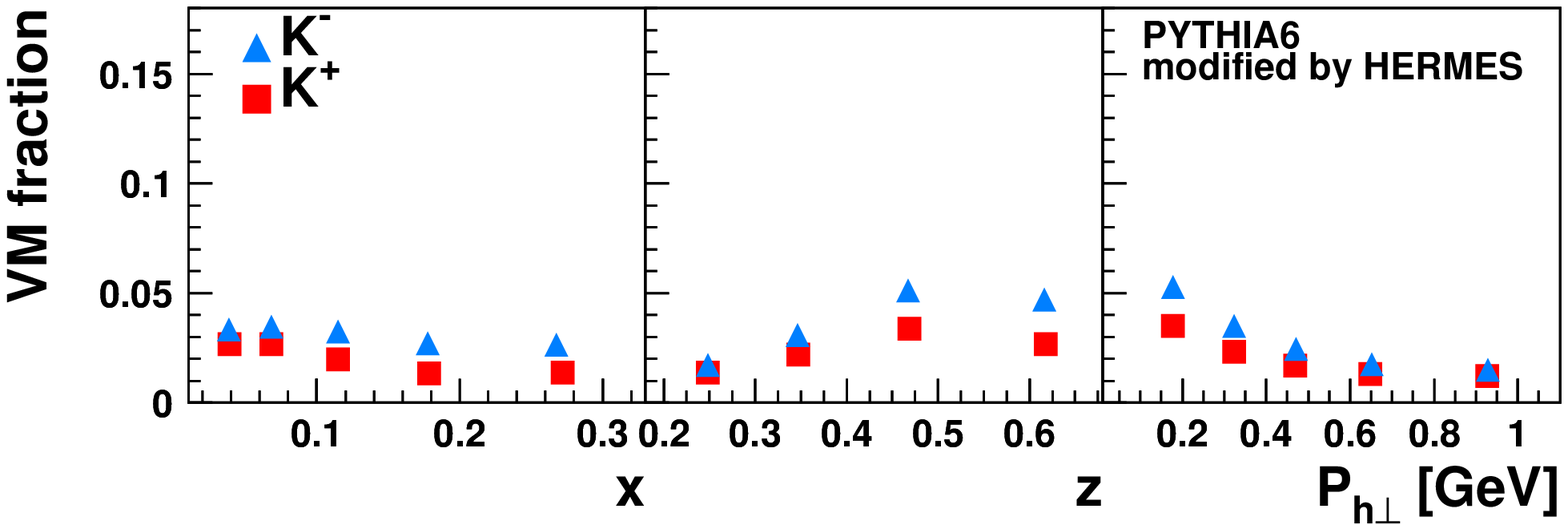}
\end{tabular}
\caption{Collins amplitudes (upper panel) and Sivers amplitudes
(middle panel) for both charged pions (left column, charge as
labelled) and charged kaons (right column, charge as labelled) as a
function of \(x\), \(z\) and \(\boldsymbol{P}_{\,\text{h}\perp}\). The
error bands represent the maximal systematic uncertainty due to hadron
misidentification, due to acceptance and detector smearing effects and
due to a possible contribution from the
\(\left<\cos{\phi}\right>_{\text{UU}}\) amplitude in the
spin-independent cross section. The common overall \(6.6\%\) scaling
uncertainty is due to the target polarisation uncertainty. The lower
panel shows the fraction of charged pions (left column) and charged
kaons (right column) produced in vector meson decay as simulated by
a version of \textsc{PHYTHIA6} \cite{Sjostrand:2000wi} tuned for
\textsc{HERMES} kinematics.}
\label{ssa-amplitudes}
\end{figure}

The average Collins amplitude is positive for \(\pi^+\) and negative
for \(\pi^-\). Also, the magnitude of the \(\pi^-\) amplitude appears
to be comparable or larger than the one for \(\pi^+\). One explanation
could be a substantial magnitude for the disfavoured Collins
fragmentation function with an opposite sign than the favoured Collins
fragmentation function. For charged kaons no significant non-zero
Collins amplitudes are found. However, the Collins amplitudes for
\(K^+\) are consistent to the \(\pi^+\) amplitudes.

The significantly positive average Sivers amplitudes observed for both
\(\pi^+\) and \(K^+\) imply a non-vanishing orbital angular momentum
of the quarks inside the nucleon. For \(\pi^-\) and \(K^-\) the
averaged Sivers amplitudes are consistent with zero.

\begin{theacknowledgments}
This work has been supported by the German Bundesministerium f\"ur
Bildung und Forschung (\textsc{BMBF}) (contract nr.~06 ER 125I) and
the European Community-Research Infrastructure Activity under the FP6
''Structuring the European Research Area'' program (HadronPhysics I3,
contract nr.~RII3-CT-2004-506078).
\end{theacknowledgments}

\bibliographystyle{aipproc}

\bibliography{spin-proceedings-diefenthaler}

\end{document}